\pdfoutput=1
\documentclass[12pt]{article}        

\usepackage{amsmath}
\usepackage{amssymb}
\usepackage{graphicx}
\usepackage{cite}
\usepackage{url}
\begin{document}

\begin{titlepage}

\vspace{-20mm}
\begin{flushright} 
{\bf IFJPAN-IV-2010-5}\\ 
{\bf MCNET-10-12}\\ 
\end{flushright}
\vspace{2mm}
\begin{center}

{\Large\bf
  MCdevelop -- the universal framework for Stochastic Simulations\footnote{
This work is partly supported by by the European Community's Human Potential
Programme ``Doctus", the Marie Curie research training network ``MCnet'' 
(contract number MRTN-CT-2006-035606) and the Polish Ministry of 
Science and Higher Education grant  No. 1289/B/H03/2009/37.}
}
 \end{center}

\begin{center}
  \large{\bf M. Slawinska}
\end{center}
\begin{center}
{\em Institute of Nuclear Physics, Polish Academy of Sciences,\\
    ul. Radzikowskiego 152, 31-342, Krak\'ow, Poland}
\\
\end{center}

\begin{center}
  \large{\bf S. Jadach}
\end{center}
\begin{center}
{\em Institute of Nuclear Physics, Polish Academy of Sciences,\\
    ul. Radzikowskiego 152, 31-342, Krak\'ow, Poland,}
\\
{\em 
Theory Group, Physics Department, CERN, CH-1211, Geneva 23, Switzerland \\
}
\end{center}

\vspace{3mm}
\begin{abstract}
We present {\tt MCdevelop}, a universal computer framework for
developing and exploiting the wide class of Stochastic Simulations (SS) software.
This powerful universal SS software development tool
has been derived from a series of scientific projects 
for precision calculations in high energy  physics (HEP), which
feature a wide range of functionality
in the SS software needed for advanced precision Quantum Field Theory
calculations for the past LEP experiments 
and for the ongoing LHC experiments at CERN, Geneva.
 {\tt MCdevelop} is a ``spin-off'' product of HEP
to be exploited in other areas, while it will still
serve to develop new SS software for HEP experiments.
Typically SS involve independent generation of large sets of random ``events",
often requiring considerable CPU power.
Since SS jobs usually do not share  memory
it makes them  easy to parallelize.
The efficient development, testing and running
in parallel SS software requires a convenient 
framework to develop software source code,
 deploy and monitor batch jobs, merge and analyse results
from multiple parallel jobs, even before the production runs are terminated. 
Throughout the  years of development of stochastic simulations for HEP,
a sophisticated framework featuring all the 
above mentioned functionality has been implemented.
{\tt MCdevelop} represents its latest version,
written mostly in C++ (GNU compiler gcc).
It uses {\tt Autotools} to build binaries
(optionally managed within the {\tt KDevelop} 3.5.3 
Integrated Development Environment (IDE) ).
It uses the open-source ROOT package
for histogramming, graphics and
the mechanism of persistency for the C++ objects.
{\tt MCdevelop} helps to run multiple parallel jobs
on any computer cluster with NQS-type batch system.
\end{abstract}

\vspace{2mm}


\end{titlepage}

\noindent{\bf PROGRAM SUMMARY}
\vspace{10pt}

\noindent{\sl Title of the program:}\\
 {\tt MCdevelop}.

\noindent{\sl Computer:}\\
Any computer system or cluster with C++ compiler and 
UNIX-like operating system.\\

\noindent{\sl Operating system:}\\
most UNIX systems, Linux\\
The application programs were thoroughly tested 
under Ubuntu 7.04, 8.04 and CERN Scientific Linux 5.

\noindent{\sl Programming languages used:}\\
ANSI C++

\noindent{Other requirements:}\\
(i) ROOT 
    package installed, version 5.0 or higher.\\
(ii) GNU compiler gcc and GNU Build System {\tt Autotools} --
optionally within {\tt KDevelop} 3.5.3 integrated development environment.
\\
(iii) NQS-type batch system (For running jobs in a parallel mode)

\newpage

\tableofcontents

\newpage

\section{Introduction}

Multidimensional generation/integration is an important ingredient in a 
variety of computational problems in science, finance and industry. 
The common and efficient techniques employed in these problems are 
Stochastic Simulations (SS), known also as Monte Carlo (MC) methods. 
These algorithms often require large CPU power,
but can be easily parallelized  and run on multi-node computer clusters,
also referred to as PC-farms.

Stochastic Simulations have been applied in simulations
in high energy physics (HEP) since the early 1960's.
From its early beginnings SS programs were accompanied with an auxiliary library
of random number generators, histogramming, job control scripts to compile/run etc.
With the advent of PC-farms run under UNIX in the early 1990's, the
system environment of SS programs was supplemented with {\tt makefile} scripts,
batch system scripts for running jobs in parallel,
software tools for collecting results from many jobs,
more sophisticated management of input/output files
(parsers to read input data, a primitive form of persistency, output compression).
The last decade has added to the above arsenal of tools the use of the fully
object oriented programming language C++ and IDEs such as
{\tt Eclipse} or {\tt KDevelop}.

The state of the art in these auxiliary system tools for SS computations
of the late 1990's is well represented by these included
in the SS/MC project {\tt BHLUMI}~\cite{bhlumi4:1996}, which was
dedicated to calculations of quantum electrodynamical effects in the
small angle electron-positron scattering at the LEP experiments.

The present {\tt MCdevelop} system owes a lot to the above project.
The actual event generator BHLUMI and its
all auxiliary programs were written in FORTRAN77.
It included a library of random number generators, the
simple but powerful histogramming package {\tt glibk},
featuring LaTeX-based graphics and a primitive form of persistency,
that is a possibility of dumping into a disk file the status of the MC generator
and all histograms for the purpose of resuming the MC production later on.
Multipurpose use of the BHLUMI generator consisting of 3 subgenerators
was managed by a custom set of interconnected makefiles in many subdirectories.
This auxiliary part of BHLUMI also included system of 
{\tt makefile} and {\tt csh} shell scripts managing multiple parallel
jobs on an early PC-farm under the NQS batch system.
The above was representative of the state of art in the 1990's and most of its
functionality is preserved in the present {\tt MCdevelop} system.

Over the last decades the above multifunctional auxiliary system derived from BHLUMI
was ported to C++, the system of {\tt makefiles} is no longer written by hand
but rather managed semi-automatically by {\tt Automake}
(one of the tools in GNU Build system {\tt Autotools}).
Moreover, a modern integrated development environment (IDE) {\tt KDevelop}
was introduced into the everyday source code development process.
Histogramming and graphics are now done with help of the ROOT library.
This new incarnation of the older system
has been already employed in the development of various
projects related to LHC physics 
(for instance CMC~\cite{Jadach:2007qa} or EvolFMC~\cite{Jadach:2008nu})
and will be used
for developing other SS/MC projects for HEP in the future.
We hope that other developers of SS software, also beyond HEP, will also
profit from its great efficiency and rich functionality.

Let us list complete specification of the functionality of
the SS software development and execution framework, as implemented in {\tt MCdevelop}:
\begin{enumerate}
\item
Generally, it should be universal, that is easily adjustable for any
SS application of the interest.
\item
Reduced dependence on non-open source codes and proprietary libraries,
without sacrificing required functionality.
\item
Histogramming, graphics and random number generators
from external open-source packages integrated with the main code.
\item
Availability of the modern source code tools such as a syntax-aware
editor and visualisation of the object classes.
\item
Easy semi-automatic configuration of the compilation/linking parameters,
paths to system libraries, environmental variables,
easy access to debuggers (including debugging of memory leaks), etc.
\item
The framework should facilitate off-line analysis of results
with help of suitable graphical libraries and convenient I/O methods,
in particular the use of persistency mechanism of the C++ objects
should be implemented and fully exploited.
\item
Transparent and flexible methodology of setting up
input data and other configuration parameters,
both in single-node and PC-farm execution mode.
\item
Capability of running easily multiple jobs on a PC-farm under the NQS-like batch system --
it should be easy to switch from one-node to multi-node execution mode,
without any modifications to the code and input data.
\item
Deployment of multiple jobs on the PC-farm should
include setting up separate working directories for each job
(with different random number seed initialisation for each of them),
easy starting and stopping all jobs,
combining output (histograms) from multiple output files in all working
directories into a single output files, etc.
\item
While running on a PC-farm, one should be able to
do on-line analysis of the partial accumulated results,
that is to inspect these partial MC results
(combined from all running jobs)
without stopping the production of the PC-farm.
\end{enumerate}

The framework we present here has all the above listed features.
The particular solutions implementing basic functionality will be 
presented in Section 2 and the use of computer farm within {\tt MCdevelop} 
framework  in Section 3.
The external package chosen to supplement the functionality of {\tt MCdevelop}
is  ROOT~\cite{Brun:1997pa}. ROOT provides histogramming, graphics, and
persistency of the C++ objects, as well as the random number generator {\tt Foam}.
Moreover, the optional use
of KDevelop 3.5.3, the  IDE of the popular desktop environment KDE (www.kde.org)
provides integrated source code development and testing 
environment including runtime debuggers.
It is conveniently integrated with {\tt Autotools}, which we use for configuration,
compilation and linking.

The use of the powerful
general purpose simulation tool  {\tt Foam}~\cite{foam:2002}
within {\tt MCdevelop}
is very easy due to its inclusion in ROOT.
For the users interested mainly in the use of {\tt Foam} 
{\tt MCdevelop} provides a convenient integrated working environment.

Let us elaborate a little bit more on the role of {\tt Foam}.
With the presently available CPU power it is possible and convenient
to use some general purpose tool for generating or integrating an arbitrarily
complicated multidimensional distribution, typically up to dimension $n\sim 15$,
instead of inventing custom Monte Carlo algorithm adjusted to
particular shape of the distribution (integrand), as it was
recommended two decades ago.
Object(s) of the class {\tt TFoam} may be useful in any kind and size of a SS
project;
in the smaller one it may actually be an essential part of it,
while in the big one may serve as a component(s).
{\tt Foam} is primarily aimed to generate automatically
MC events with unit weight according to an arbitrary multidimensional
distribution provided by the user.
It can also be used for numerical integration%
\footnote{For non-positive distribution integration is done using
weighted MC events.}.
Foam works in two stages: (i) initialisation, in which it divides
the integration domain into system cells, in such a way that cells are
smaller and cover more densely the regions where the user distribution
varies strongly (has peaks)
(ii) generation, when it generates MC events {\em exactly} according
to the user (pre-)defined  distribution	.
The user may request for either weight one events or weighted events.
{\tt MCdevelop} facilitates the use of {\tt Foam} -- in particular
it provides an {\em interface} to the user distribution function
in its base class.
We do not elaborate in this document on the use of {\tt Foam},
its steering parameters etc.,
as it is accompanied with its own detailed user 
manual~\cite{foam:2002} and there are several examples of its use
in the subdirectory {\tt tutorials} of the ROOT distribution, see
\url{http://root.cern.ch/root/html/tutorials/foam/index.html}.

Let us remark that every HEP experiment has its own software
environment for running massive production (simulation)
of the Monte Carlo events on PC-farms
with the functionality similar to that of {\tt MCdevelop}.
The main difference between them is that the main
aim of {\tt MCdevelop} is to develop source code
of the MC event generators and testing them extensively,
while HEP experiments rather concentrate on their use.
Also, HEP experiments store MC events on disks, while {\tt MCdevelop}
is oriented towards booking and filling many histograms.

In the next Section we describe how the above non-trivial goals were achieved.
In fact they determine quite rigidly the organisation and structure
of {\tt MCdevelop}.
We briefly present the workflow of the program,
functionality and structure of its main C++ classes.
Section 3 is the user manual describing its execution
on the different levels of proficiency.
The more technical details extending the functionality of {\tt MCdevelop}
framework, such as the use of {\tt Autotools}
and {\tt KDevelop} IDE will be explained in Sections: \ref{sec:Autotools}
and \ref{sec:KDevelop} in particular in the context of linking with (external) ROOT~\cite{Brun:1997pa} libraries.
%

\section{Structure and functionality of the code}

{\tt MCdevelop} was derived from the existing SS projects such as
BHLUMI~\cite{bhlumi4:1996} and EvolFMC~\cite{Jadach:2008nu},
by means of extracting (abstracting) their universal part, such that
it can be easily exploited in other SS applications, not only within HEP.
A transparent, modular structure is achieved by means of design
and implementation of its C++ classes,
organisation of the data flow reflecting the needs of typical SS project,
and its directory structure.

In this Section we will describe, how the framework is split into a
main library of C++ programs and template subprojects, and discuss 
how they are interrelated. 
The workflow of the typical MC production run will be described.
Main C++ classes  will be described in some details at the end of this section.

\subsection{The distribution directory}

{\tt MCdevelop} is located in a single UNIX directory consisting of
the following subdirectories:
\begin{itemize}
\item {\tt MCdev} 
- contains universal part of the framework,
  library of base classes and scripts for
  setting up and running multiple jobs in parallel on a PC-farm;
\item {\tt Template0} and {\tt Template} 
- collect demonstration applications,
  templates of user SS applications;
\item {\tt m4} 
- required by {\tt Automake} in order to link the project
  with external libraries.
\end{itemize}
\begin{figure}[!ht]
\begin{center}
\includegraphics[width=7cm]{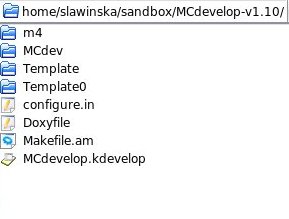}%
\caption{The list of files and subdirectories in the main directory of {\tt MCdevelop}.}
\end{center}
\end{figure}

\subsubsection{Content of the main subdirectory {\tt MCdev}}
\label{sect:mcdev}

The {\tt MCdev} subdirectory contains the main part of {\tt MCdevelop} 
source code: headers and implementations of the base classes 
{\tt TMCgen}, {\tt TRobol} and a few auxiliary classes,
which are used to build and link the main library {\tt libMCdev}.
Most of the classes in the SS project developed under {\tt MCdevelop}
are derived (inherited) from an appropriate base class mentioned above.
The subdirectory {\tt MCdev/farming} contains also several scripts
(interpreted ROOT macros), which
can be used to setup parallel batch jobs on a PC-farm with the NQS batch system,
submit them, monitor their performance and optionally stop them.
Finally, they help to combine output results (histograms)
from any number of working directories into a single output file.

\begin{figure}[!h]
\begin{center}
\includegraphics[width = 7cm, height=10cm]{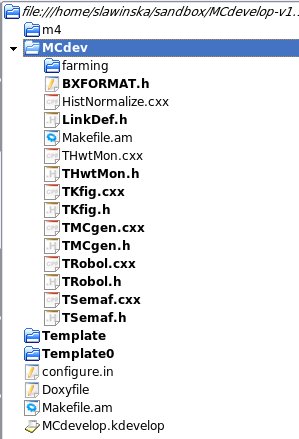}%
\caption{List of files and directories in the main subdirectory {\tt MCdev}.}
\end{center}
\end{figure}
\subsubsection{Two templates of the user projects}
\label{sect:templates}

The distribution directory of {\tt MCdevelop} contains not only
the core framework but also two templates of the SS projects
based on {\tt MCdevelop}.
Their role is to guide potential users in using {\tt MCdevelop} 
or customising their own SS project within the {\tt MCdevelop} framework.
They may also be exploited as a starting point to develop the user's own
SS project from the scratch within the {\tt MCdevelop} framework.

These template projects can be found in subdirectories {\tt Template0} and {\tt Template}.
Their structure is typical of projects already developed within the {\tt MCdevelop} framework.
Each of these demonstration project contains the main execution program {\tt MainPr}
which generates a series of MC events,
and a program {\tt XPlot} analysing results of SS/MC run.
{\tt XPlot} exploits the graphics capabilities of ROOT.
The simulation run is performed in the subdirectory {\tt work}
where an initialisation script {\tt Start.C}  and all output files are placed.
{\tt Start.C} is a small C++ macro interpreted by ROOT, enabling the user to 
initialise adjustable parameters of the MC run (eg. number of events to generate).
Of course, in a bigger SS projects there will be several
analysis programs like {\tt XPlot}, possibly in a separate subdirectory,
and/or several subdirectories like {\tt work}.

\begin{figure}[!ht]
\begin{center}
\includegraphics[width = 7cm, height=10cm]{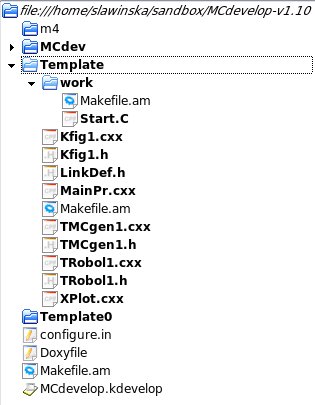}%
\caption{Files in a Template subdirectory }
\end{center}
\end{figure}

\subsection{Persistency with the help of ROOT}
\label{subsect:persistency}

Persistency is an additional feature in the object oriented programming (OOP)
framework, which allows to write an entire object into disk file,
then read it from the disk later on by the same job
or other job in the ``ready to go'' state.
From our past experience we know, that the typical SS software project
gains in functionality through the persistency of the following objects:
(i) the random number generator common to all components of the MC generator,
(ii) the entire MC generator, object of a single MC event,
(iii) the module analysing  well defined aspects of the simulation
during the execution.

C++ does not support persistency,
hence it has to be added with help of external software tools.
In our case we profit from the implementation of persistency 
in the ROOT~\cite{Brun:1997pa} package.
Let us remark that any of the tools adding persistency on top of C++
has to deal with the difficult problem of storing and regenerating pointers
inside a single object and among interrelated objects.
The solution of this problem in ROOT is working well, however it requires
special care in the implementation (special directives in the headers).

We characterise the important role of persistency in the functionality of
any SS project  in Section~\ref{sect:workflow}.
Here, we will briefly explain the importance of persistency on the example
of the random number generator and a ``semaphore" object.

The object of the central random number generator
is provided by one of ROOT classes.
It is allocated inside a configuration script {\tt Start.C}
(see below for more details about this script).
After (optional) resetting of its
initial seed it is immediately written into the disk file 
(by default {\tt mcgen.root} ).
At the start of the production it is restored from the disk and a pointer to it
is distributed all over components of the Monte Carlo event generator.
It is important that this object is the same and only one random number generator object
for the entire MC generator. For a parallel execution 
it is cloned from the disk file for each batch job,
and its random number seed is reinitialised ``in the flight'', such that different parallel batch jobs use different random number series from the same instance of 
the r.n. generator.
At the end of the MC run the object of r.n. generator is dumped into the disk
and can be easily read from the disk in the case of continuing MC run in the next
batch job, as if there was no break in the MC generation run at all.

The object of the MC event generator class (inheriting from the
base class {\tt TMCgen}) undergoes similar history as r.n. generator described earlier.
The important difference is that it may contain many objects
of other classes, including {\tt Foam}, or pointers to these objects.
Its initialisation consumes often considerable CPU time
and is performed in several steps. 
It is therefore quite profitable to be able to write such an operational
object of the MC generator, or several version of it
(for instance initialised with different input parameters)
into a disk file for the later use.
Let us stress that ROOT is capable to write/read into diskfile
an entire object which consists of many other objects,
even the ones referred to by pointers.

The other persistent object implemented and used in {\tt MCdevelop} 
during the execution of the program,
is a special small auxiliary object
of the {\tt TSemaf} class. Its role is to
control the execution of multiple batch jobs on PC-farm.
This and other objects are read/written from/to disk files many times
during MC production run,
see Sections~\ref{sect:workflow} and~\ref{sect:run} for more details.


\subsection{The workflow }
\label{sect:workflow}

Before moving to implementation details,
let us overview the general workflow of the program.
This will help to better understand the functionality 
and data structure of the classes as well as the critical role of persistency
in their implementation.
The role and interrelations of the main components presented in the previous Sections ~\ref{sect:mcdev} and ~\ref{sect:templates} 
and classes described in next Section \ref{sect:classes}
will then become clearer.

The two basic components in the workflow are:
the interpreted C++ script {\tt Start.C}
and compiled C++ program {\tt MainPr}.
{\tt MainPr} is located in the top subdirectory of a given (sub)project
and by default is identical for all (sub)projects. It is the main, 
universal production executable.
{\tt Start.C} is placed in the subdirectory {\tt work} and is specific for 
a project. It is already called before MC production. 
The purpose of  {\tt Start.C} is to allocate objects of
(i) the random number generator of the {\tt TRandom} base class,
(ii) the MC event generator of the {\tt TMCgen} base class
and (iii) objects of the analysis module of the {\tt TRobol} class.
Moreover  {\tt Start.C} is the place where user may easily customise
these objects by resetting their default parameters (data members).
Typically, the user may reset in {\tt Start.C}
random number seed and any configuration/input parameters
in the {\tt TMCgen} object before the actual initialisation.
(It is good practise to set these parameters in the constructor
to some default values and optionally reset them in {\tt Start.C}.)

{\tt Start.C} also allocates a small object of the  {\tt TSemaf} class
used to control execution of the main loop over MC events in 
the main program {\tt MainPr}.
Objects of the MC generator originally allocated in {\tt Start.C}
is in the ``preliminary form'', before any genuine initialisation.
The instances of the all above objects are then saved into disk
files created at the end of {\tt Start.C} execution.
They are {\tt semafor.root} and {\tt mcgen.root} files
encoded in the ROOT format (with compression).
Note that all the above objects at the time of their
allocation and customisation in {\tt Start.C}
are at this stage not interrelated,
for instance with the help of pointers.

It is the role of {\tt MainPr} to read the object of
the Monte Carlo event generator and other related
objects from disk files  {\tt mcgen.root}, {\tt histo.root},
{\tt semafor.root} (created by {\tt Start.C}),
to assemble/initialise the working object of the MC event generator
and to run the main loop generating series of the MC events.

We follow the policy of keeping the part of software managing the MC production, collecting and analysing
average quantities and distributions all over the entire series of the MC events
well separated from the MC event generator.
In {\tt MCdevelop} this analysis role is reserved for an objects of the classes
inheriting from the persistent base class {\tt TRobol}.

During the generation of a long series of the MC events
in {\tt MainPr}, the object of the {\tt TRobol} class is invoked
to analyse each MC event and accumulate all interesting information
in the histograms.
After generating a well defined subsample series of the MC events,
which we shall refer to as an {\em event group},
(the number of MC events in the group is defined by the user)
{\tt MainPr} dumps the actual copy (status) of the MC event generator object
and the object of {\tt TRobol} class containing all histograms into 
{\tt mcgen.root}, {\tt histo.root} files.
In principle these objects are in the ``ready to go'' state.
If the user wishes to restart MC generation in the next production job,
they will be fully functional, as if there was no break in the production --
without any need of the re-initialisation of the MC generator object.

After generating each event group of the MC events,
{\tt MainPr} is reading object of the {\tt TSemaf} class from
{\tt semafor.root} file.
This object contains the text flag defining the state of MC production:
``START'', ``CONTINUE'', or ``STOP''.
The initial value of the flag in {\tt Start.C} is ``START''.
It is changed in {\tt MainPr} to ``CONTINUE''.
However, the user has the possibility to change this flag in the object
in {\tt semafor.root} file to ``STOP'', even before the end of the job.
Once {\tt MainPr} discovers ``STOP'', it terminates the loop over MC events.
The above method provides protection against
unexpected code or machine crashes,
because only a fraction of results are lost.
The saved state of MC generator can be used not only to  
continue or resume Monte Carlo production, and is also useful for debugging
program crashes after generating long series of MC events.

Obviously, for the above workflow the extensive use
of persistency is critical and instrumental.

Finally, let us note that in the file {\tt histo.root}
distribution histograms as {\tt TH1D} or {\tt TH2D} {ROOT} objects
are stored not as data members of the {\tt TRobol} object,
or using one of the container classes of ROOT,
but rather as a loose collection of histograms 
accessed with help of text type ``keys''.
This organisation is chosen mainly for historical reasons
and can be replaced by something more sophisticated.

\subsection{Structure of the source code}
\label{sect:classes}

\begin{figure}[!ht]
\centering
\includegraphics[width = 7cm, height=10cm]{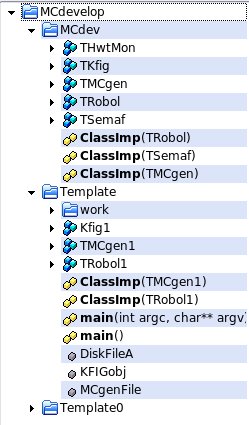}
\caption{
   C++ classes of {\tt MCdevelop} and their location.
   Base classes are placed in {\tt MCdev} subdirectory
   and derived classes are located
   in the subdirectory of the user example project {\tt Template}.}
\label{fig:classes}
\end{figure}

{\tt MCdevelop} consists of several base classes
located in subdirectory {\tt MCdev} and derived classes,
specific for each subproject,
located in the subdirectory of a given subproject.
This is shown in Figure~\ref{fig:classes}.
Each (sub)project directory has its own set of dedicated classes 
derived from the base classes and builds its own project's library.
Main base classes of {\tt MCdevelop} 
are also listed in Table~\ref{tab:classes}.
In the following we shall describe the base classes in a more detail.

\begin{table}[!ht]
\centering
\begin{small}
\begin{tabular}{|l|p{14.0cm}|}
\hline
{\bf Class} & {\bf Short description}\\ 
\hline\hline
\multicolumn{ 2}{|c|}{{ Main classes: } }\\
\hline
\hline
{\tt TMCgen} &
Abstract base class  for any type of MC generator. 
It handles generation of MC events according to a user-defined distribution.
The method {\tt Generate()} is virtual and needs to be implemented
in derived classes.\\
\hline
{\tt TRobol} &
An object of this class handles MC production, collects and saves 
the raw material for analysing MC results at the end of the MC run.
It features a three-stroke engine of the methods for analysis:
Initialise-Production-Finalise. Its virtual functions must be 
reimplemented by the user in derived classes.\\
\hline\hline
\multicolumn{ 2}{|c|}{{ Auxiliary classes: } }\\
\hline\hline
{\tt TSemaf} &
An object of this class helps to manage/control the MC event production loop
in the main program {\tt MainPr}.
Main program reads from this object a semaphore variable
{\tt m\_flag} in order to decide whether to terminate or continue job execution.\\
\hline
{\tt TKfig} &
Library of auxiliary procedures for digesting the MC results.
It includes procedures for normalising semi-automatically
all the histograms at the end of the MC run.
Its functions {\tt DefHist1D()} and  {\tt DefHist2D()}
can be extended/customised in order
to compare histograms from the MC run with the analytical distributions.\\
\hline
\hline
\end{tabular}
\end{small}
\caption{ Description of C++ base classes of {\tt MCdevelop}.}
  \label{tab:classes}
\end{table}

\subsubsection{The generator class {\tt TMCgen}}

The {\tt TMCgen} class is the base class from which the actual MC generator
class should be derived.
Although the use of {\tt Foam} in the MC generator is not mandatory,
we assume that it will be used quite often, hence we facilitate it by means
of the inheritance from the interface class {\tt TFoamIntegrand}.
In this way  {\tt TMCgen} may implement the {\tt Density} method
providing the distribution to be generated by {\tt Foam}.\\
The constructor of the class
{\tt TMCgen(const char* Name)} creates the TMCgen object,
whose name is specified in the argument.
For example, the instruction:
\\
{\tt TMCgen * MCgenerator = {\bf new} TMCgen("MCgenerator")};
\\
creates an instance of {\tt TMCgen} generator named {\tt MCgenerator}.
The generator instance is originally allocated in the {\tt Start.C} script.
There the user can adjust some of the parameters.
A typical change is in the generated distribution,
which can be done in the following:
\\
{\tt
MCgenerator->m\_MEtype = "Example"; //Matrix Element type
}\\
The complete initialisation of the object
of this class is done automatically by {\tt MainPr}.
The main data members of the {\tt TMCgen} class are summarised
in Table~\ref{tab:MCgen_members}.

\begin{table}[!h]
\centering
\begin{small}
\begin{tabular}{|l|p{11.0cm}|}
\hline
{\bf {\tt TMCgen} data member} & {\bf Short description},\\ 
\hline
\hline
{\tt char f\_Name[64] } &
Name of a given instance of {\tt TMCgen} class,\\
{\tt   float f\_Version} &
actual version of the program,\\
{\tt   char  f\_Date[40]} &
release date of {\tt MCdevelop}.\\
\hline
{\tt TRandom  *f\_RNgen} &
  External random number event generator,\\
{\tt  TFoam    *f\_FoamI} &
  the object of the {\tt mFoam} class for generating the user provided density distribution,\\
{\tt  TH1D     *f\_TMCgen\_NORMA} &
 special ROOT histogram keeping overall normalisation.\\
\hline
{\tt  int f\_IsInitialized} & 
A flag  to prevent repeating initialisation of an instance during the MC run,\\
{\tt double    f\_NevGen} &       
event serial number,\\
{\tt  ofstream *f\_Out}$^s$ &
 external logfile for messages, \\
{\tt  TString m\_MEtype}&
the type of distribution to be generated.\\
\hline
\hline
\end{tabular}
\end{small}
\caption{
 Data members of {\tt TMCgen} class. Members indicated with superscripts $s$
 are provided with streamers.}
\label{tab:MCgen_members}
\end{table}

In Table~\ref{tab:MCgen_functions} we describe the main functions/methods
of the {\tt TMCgen} class.
Among them are {\tt Initialize}, {\tt Generate} and {\tt Finalize},
called at the relevant stages of the Stochastic Simulation code.

The {\tt Density} function provides a density distribution to be generated
by {\tt Foam}.
In practice the distribution generated by the MC generator object
may have several variants,
hence the presence of the flag {\tt m\_MEtype}, which can be reset as follows:\\
{\tt
 MCgenerator->m\_MEtype = "Example";
 }\\
{\tt Density} uses two arguments:
dimensionality {\tt kDim} of a distribution to be generated
and an input vector/point provided by {\tt Foam}.
{\tt Density} returns the value of the distribution
to be generated at this input point.
{\tt Foam} keeps track of the normalisation of the distribution of {\tt Density},
that is provides the integral over this distribution at the end of the MC run.
{\tt Foam} may generate weighted or unweighted MC events, see manual of 
{\tt Foam}~\cite{foam:2002} for more information.

\begin{table}[!h]
\centering
\begin{small}
\begin{tabular}{|l|p{6.0cm}|}
\hline
{\bf {\tt TMCgen} function} & {\bf Short description}\\ 
\hline\hline
\multicolumn{ 2}{|c|}{{ Constructors and destructors: } }\\
\hline
\hline
{\tt TMCgen()} &
 Explicit default constructor for\\ & ROOT streamer,\\
{\tt  TMCgen(const char*)} &
 user constructor,\\
{\tt  \~~TMCgen()}&
explicit user destructor.\\
\hline\hline
\multicolumn{ 2}{|c|}{{ Main methods: } }\\
\hline
\hline
{\tt  virtual void} 
& Initialisation of main members \\
{\tt $\quad$ Initialize(TRandom*, ofstream*, TH1D*)} & and allocation of memory.\\
{\tt  virtual void } & For optionally correcting pointers \\ 
{\tt $\quad$ Redress(   TRandom*, ofstream*, TH1D*)}& reconstructed by streamers,\\
 & not implemented in the base class.\\
{\tt  virtual void Generate()} &
Generate a single MC event, \\ & not implemented in the base class.\\
{\tt  virtual void Finalize()} &
Finalise MC  run, produce \\ & final printouts.\\
{\tt  virtual double }  & Generates a single point from \\ 
{\tt $\quad$ Density(int nDim, double *Xarg){;} } & a given density distribution; \\ 
 & not implemented in the base class\\
\hline
\hline
\end{tabular}
\end{small}
\caption{ Methods of {\tt TMCgen} class.}
  \label{tab:MCgen_functions}
\end{table}

\subsubsection{The MC run and analysis module class {\tt TRobol}}

\begin{table}[!ht]
\centering
\begin{small}
\begin{tabular}{|l|p{9.0cm}|}
\hline
{\bf {\tt TRobol} data member} & {\bf Short description}\\ 
\hline
\hline
{\tt char f\_Name[64] } &
Name of a given instance of the {\tt TRobol} class.\\
{\tt   double f\_NevGen} &
A serial number for each generated event.\\
{\tt   double f\_count1} &
Auxiliary event counter (used mainly for debugging).\\
{\tt   long f\_isNewRun} &
A flag which takes values from the\\
&  MC Generator method {\tt GetIsNewRun()}.\\
\hline
{\tt TRandom *f\_RNgen}$^s$ &
Central random number generator,\\
{\tt TMCgen *f\_MCgen}$^s$ &
SS (Monte Carlo) event generator,\\
{\tt TFile *f\_GenFile}$^s$ &
ROOT file with {\tt TMCgen} object,\\
{\tt TFile *f\_HstFile}$^s$ &
ROOT file with all generated histograms saved  as\\
& ROOT {\tt TH1D} or {\tt TH2D} objects,\\ 
{\tt std::ofstream *f\_Out}$^s$ &
central log file for messages,\\
{\tt std::ofstream *f\_TraceFile}$^s$& 
auxiliary log file for debugging\\
\hline
\hline
\end{tabular}
\end{small}
\caption{ Members of the {\tt TRobol} class. Members indicated with superscripts $s$
 are provided with streamers.}
  \label{tab:Robol_members}
\end{table}

\begin{table}[!h]
\centering
\begin{small}
\begin{tabular}{|l|p{6.3cm}|}
\hline
{\bf {\tt TRobol} function} & {\bf Short description}\\ 
\hline\hline
\multicolumn{ 2}{|c|}{{ Constructors and destructors: } }\\
\hline
\hline
{\tt TRobol()} &
 Explicit default constructor for\\ & ROOT streamer,\\
{\tt  TRobol(const char*)} &
 user constructor\\
{\tt  \~~TRobol()}&
explicit destructor,\\
\hline\hline
\multicolumn{ 2}{|c|}{{ Main methods: } }\\
\hline
\hline
{\tt  virtual void } & Resets all pointers after recreating \\
{\tt $\quad$ Initialize(ofstream*, TFile*, TFile*)} & objects from disk files,\\
{\tt  virtual void Production(double\&)}& 
the main function steering the SS; \\
& invokes {\tt TMCgen::Generate()}. \\
{\tt  virtual void FileDump()}& 
saves all objects into relevant files,\\
{\tt  virtual void Finalize()}& 
finalises SS and does final printouts.\\
\hline
\hline
\end{tabular}
\end{small}
\caption{ Methods of {\tt TRobol} class.}
  \label{tab:Robol_functions}
\end{table}

The main members of {\tt TRobol} class are listed in the Table~\ref{tab:Robol_members}.
The object of the {\tt TRobol} class owns pointers to random number generator,
MC generator and to all output disk files.
It is an important object in the Stochastic Simulation project code.
Its main methods are listed in Table~\ref{tab:Robol_functions}.

The {\tt Generate()} method of the MC event generator object is invoked 
in the {\tt Production()} function of the {\tt TRobol} object.
This arrangement is convenient for present applications but not mandatory.
This call could be placed in {\tt MainPr} main program.
It might be more convenient/transparent option within
a bigger SS project with several objects/classes derived from {\tt TRobol} class
dedicated to different types of analysis of the MC results.

\newpage
\section{Running stochastic simulation within the {\tt MCdevelop} framework}
\label{sect:run}

In this Section we explain how to build
the whole {\tt MCdevelop} framework,
that is configure it, compile and link with shared libraries.
The same must be also done by any SS project constructed and managed
under {\tt MCdevelop}.

In the following subsections we will describe how to execute 
two example demonstration SS projects included in the distribution.
The users of {\tt MCdevelop} will be also advised
how to develop their own new SS project with the help of {\tt MCdevelop}
and how to run massive stochastic simulations on a PC-farm.

Let us start with building {\tt MCdevelop} from the source code.
After de-archiving a copy of the {\tt MCdevelop} distribution directory
into a local directory {\tt \$MCDEVPATH},
one should type in the command line:\\
\\
{\tt \$ cd \$MCDEVPATH}
\\
{\tt \$ autoreconf -i --force}
\\
\\
After that one should execute {\tt configure.in} script:
\\
\\
{\tt \$ ./configure}
\\
\\
After running {\tt configure} script without error messages, one may execute:
\\
\\
{\tt \$ make} \\
{\tt \$ make install}
\\
\\
in order to compile and link the whole project.
By default libraries are installed in {\tt MCDEVPATH/lib}, and header fies are copied 
to {\tt MCDEVPATH/include}.

Alternatively, one can skip these two lines and comply with  instructions in the next subsection.

\subsection{Example 1 -- for an impatient user (5 min. time)}

The {\tt MCdevelop} distribution directory includes
example of a simple MC program to be run immediately after installation
is completed.

Once the framework is configured/built on the computer
(see previous section), the user may type in the command line:
\\
{\tt \$ cd Template0/work}
\\
{\tt \$ make start}
\\
The latter command invokes {\tt make install} command in {\tt MCdev} and 
{\tt Template0}, which  compiles base classes into the {\tt libMCdevlib.so} 
library and links the classes specific 
to the actual demonstration program --
the library {\tt libTemplate0.so} is built.
The C++ script {\tt Start.C} is also executed/interpreted using ROOT.
{\tt Start.C} allocates and configures a few principal objects of
the project and writes them into ROOT output files
(profiting from persistency mechanism of ROOT).
The project subdirectory {\tt Template0}
holds  copy of the main program {\tt MainPr},
which handles all MC simulation and saves results in {\tt Template0/work}.
The standard output from {\tt MainPr} will look as follows:
\\
\\
{\tt
||||||||||||||||||||||||||||||||||||||||||||||||||\\
|| MainPr:  to be generated 1000000 events\\
||||||||||||||||||||||||||||||||||||||||||||||||||\\
iEvent = 200000\\
iEvent = 400000\\
iEvent = 600000\\
iEvent = 800000\\
iEvent = 1000000\\
||||||||||||||||||||||||||||||||||||||||||||||\\
||  MainPr: Generated 1000000 events\\
||||||||||||||||||||||||||||||||||||||||||||||\\
**************************************************************\\
**************** Foam::Finalise  ************************\\
Directly from FOAM: MCresult= 1.0510578 +- 0.0001480123\\
**************************************************************\\
\\
  |--------------------|\\
  |  MainPr   Ended    |\\
  |--------------------|\\
\\
real    0m3.240s\\
user    0m2.548s\\
sys     0m0.116s\\
}

The above  SS project template generates a simple 2-dimensional
distribution using {\tt Foam}.
The program prints out the number of generated events
after generating every group of 10k MC events -- 
at the same time output the ROOT files are written and saved into the disk.
In the output we see the result of the MC integration by {\tt Foam}
together with its statistical error and the execution time of the job.
One may easily change
the number of requested total MC events in the run
and the number of events within each event group.
For this the user may correct the following lines in 
{\tt Start0.C} script in {\tt Template0/work} subdirectory:
\\ 
\\
{\tt
double nevtot   = 1e5; \\
double  nevgrp  = 2e4; }\\
\\
Optionally, in order to stop the MC run before generating all events,
one may type:
\\
\\
{\tt
\$ make stop
}\\
\\
In such a case the program will generate events until the end of the current event group,
save results and terminate.

Two dimensional histogram being
the result of the above quick run can be visualised using
small program {\tt ./XPlot0}.
To run it, one should simply type:\\
\\
{\tt
\$ cd ..\\
\$ ./XPlot0
}\\
\\
This small analysis program {\tt XPlot0} has been
already compiled/built by {\tt make} utility
simultaneously with the main program {\tt MainPr}.


\subsection{Example 2 -- running a graphical analysis program}

\begin{figure}[!h]
\centering
\includegraphics[width = 100mm]{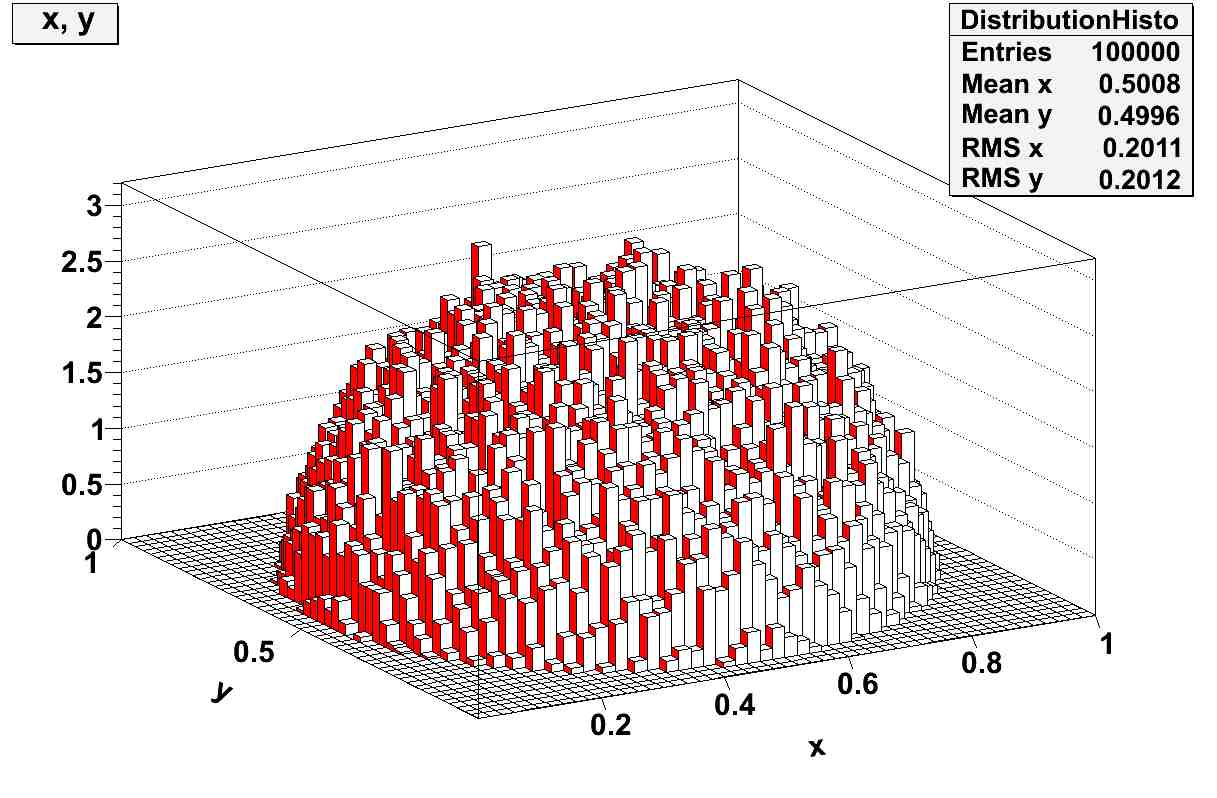}
\caption{A two-dimensional distribution generated in Template. 
   The legend summarises histogram properties.}
\label{fig:legoplot}
\end{figure}

In order to present graphically results of the MC run
one should build and execute {\tt XPlot}. It can be done
 using {\tt Automake Manager} as for {\tt MainPr}, which is
explained in detail in Section~\ref{sec:KDevelop}.
In case the  working directory path is not set, 
proceed in the same way as with {\tt MainPr}
and set the working directory path (the third line) to:
\\
\\
{\tt \$MCDEVPATH/Template/}
\\
\\
{\tt XPlot} uses {\tt MCDEVPATH/Template/work/histo.root} file 
to read out histograms as well as {\tt MCDEVPATH/Template/work/mcgen.root} 
to extract properties of the MC generator object read
from this file.
In order to use files from elsewhere the user should
edit {\tt XPlot.cxx} file or change its working directory path, 
as described above.
A typical example of analysis plot is shown in Fig.~\ref{fig:legoplot}.

\subsection{Example 3 -- running more advanced programs}

As it was already stressed, SS jobs often consume large amounts
of CPU time, hence it is profitable to run them on a PC-farm.
In the following  we will make an overview of the functionality of the
{\tt MCdevelop} framework for running jobs in parallel 
based on the existing examples.
We will also present, how our ``farming'' setup can be used in
applications.

\subsubsection{Running jobs on a computer farm}
\label{sect:farm}

We assume that NQS-like batch system is installed on the farm,
hence {\tt qsub} command and the queue class {\tt qunlimitted} 
are default parameters for submitting jobs used in the examples to follow.

Farming scripts are located in  {\tt MCdev/farming}. 
They are invoked by default from each subdirectory {\tt work} after 
configuring the project.
These scripts can help to set up a separate working directory for each batch job.
They are also able to launch batch jobs,
inspect the job's performance and merge results.

Setting up working directories is done with:\\
\\
{\tt \$ make qfarm6}
\\

The above command invokes the C++ script {\tt SetFarmQ.C}
which creates 6 working directories, 
one for each batch job, and prepares separate input files there.
The number of nodes $N$ of batch jobs can be easily adjusted to actual needs.
By default $N =6, 24, 40, 55$ are supported while executing:
\\
\\
{\tt \$ make qfarm}{\em N}
\\
\\
In order to change the number of nodes to a non-default value,
one may clone the following lines in the file
{\tt \$MCDEVPATH/Template/work/Makefile.am}:
\\
\\
{\tt
qfarm6:	farm-clean check\_all\\
\hbox{~~~~~~~~} (echo ">>>Set up MC gener:"; \$(ROOTEXEC) -b -q -l ./Start.C)\\
\hbox{~~~~~~~~} (echo ">>>Set up farm dir:"; \$(ROOTEXEC) -b -q -l \\ \hbox{~~~~~~~~~~~~~~~~}'../../MCdev/farming/SetFarmQ.C("\$(DSET)",6)' ;)\\
\hbox{~~~~~~~~}	(ln -s ../../\${MAIN} ./farm/\${DSET}.exe; echo ">>>>> DONE")\\
}\\
and set the desired number $N$ instead of 6 in the first and fourth line.\\

The next command
\\
\\
{\tt \$ make qsubmitall}
\\
\\
submits as many batch jobs,
as there are batch subdirectories made in the previous step,
with the help of the C++ script {\tt SubmFarmQ.C}.

{\tt MCdev/farming} contains additional scripts,
which enable to check the current performance, 
while running or after completing the batch jobs.
For instance:
\\
\\
{\tt  \$ make q-nev} 
\\
\\
prints numbers of generated events and status of all batch jobs in the execution.
Another command:
\\
\\
{\tt  \$ make combine}
\\
\\
merges partial results (typically 1-D and 2-D histograms)
from all working nodes and saves them to the file
{\tt histo.root} using ROOT script {\tt hadd.C}.
\\
Sometimes we want to stop all jobs immediately 
-- this can be done with the help of:
\\
\\
{\tt \$ make farm-stop}
\\
\\
Finally, removing all subdirectories created for a given series
of the batch jobs can be done using:
\\
\\
{\tt \$ make farm-clean}

\subsubsection{Configuring farming scripts for other applications}

The solution applied for parallel execution of jobs implemented within 
the {\tt MCdevelop} framework is universal and can be used in other projects, too.
Its biggest advantage is that the user's code does not need to be changed
while moving from a  sequential to parallel mode.
The required customisations of the configuration setup of {\tt MCdevelop}
for using it on any PC farm will be described in this section.

If the name of the local queue and queue class is different than default
values, one should configure the project by running:
\\
{\\ \tt
\$ ./configure --with-queue=local\_queue --with-class=local\_class
\\}\\
before proceeding further.

The scripts from {\tt MCdev/farming} are typically invoked from the
{\tt work} directories. The relevant paths are specified in {\tt work/Makefile.am}
and should be adjusted by the user for the use within a different directory structure.
All these scripts are ROOT interpreted macros written in C++,
therefore a proper installation of
the ROOT package on the PC-farm is required. The environmental variables of ROOT
are defined for the {\tt MCdevelop} framework by {\tt Autoconf} macros in {\tt \$MCDEVPATH/m4/root.m4}.

\subsection{Developing a new project in {\tt MCdevelop} framework}
\label{sect:developingnew}

The distribution directory can easily accommodate
a user's completely new SS project. Let us stress that
{\tt MCdevelop} is a convenient framework for development of both simple small
new projects similar to the ones in two demonstrations directories 
{\tt Template0} and {\tt Template} and also (in fact mainly)
for development of the large size SS project,
accompanied by many testing programs and user applications.

Simple example projects {\tt Template0} and {\tt Template}
of the distribution directory may be treated as tutorials,
and/or may be also cloned into a new directory
and customised to become a new SS project.
Let us now instruct briefly the user
how to develop a simple application in a new project's directory,
following the schemes implemented in the demonstration projects.

Any change to the directory structure must be known by {\tt Autotools},
so that  the project can be correctly built. The details of customising
the {\tt Autotools} configuration scripts are described
in Section~\ref{sec:Automake_new_project}.

While developing a new project within {\tt MCdevelop} framework, the amount
of the necessary customisations of the existing code can vary.
For instance the user may only provide his own density distribution function(s)
for {\tt Foam} for generating MC events in the MC generator
class inheriting from {\tt TMCgen}.
New histograms may be defined/added in a class deriving 
from {\tt TRobol} and new programs for graphical analysis can be  developed.

Let us stress that {\tt MCdevelop} is first of all
a convenient framework for developing
{\em large size systems} of SS programs,
in fact much bigger and more complicated
than the {\tt MCdevelop} itself and any of the example projects
included in the distribution directory.
In such a big SS project we assume that
a new sophisticated MC event generator of the user will be
developed and tested.
It will still inherit from the class {\tt TMCgen}.
However, it will be constructed using many objects of many C++ classes.
Also, typically, in such a big SS project there will be several
different {\tt TRobol} classes/objects
dedicated to specific types of testing and/or using generated MC events.
These extensions may also involve modifications
of the base {\tt TRobol} class itself,
such that its sole role is analysing MC events;
the object of the MC event generator will invoked
only in the {\tt MainPr} program,
and not in the methods of {\tt TRobol}
(as is done presently).

\section{Using {\tt Autotools} for configuring and customising the {\tt MCdevelop} framework}
\label{sec:Autotools}

The use of {\tt Autotools} within {\tt MCdevelop} plays an auxiliary, yet
important role. In this section we present basic ideas of maintaining 
project build through configuration scripts, focusing on  customisation
of this build configuration for the purpose of more advanced projects.
The reader familiar with {\tt Autotools} may skip this part of the manual.

\subsection{Extending the directory structure}
\label{sec:Automake_new_project}

Compilation and linking of {\tt MCdevelop} itself and of the SS projects
developed using  {\tt MCdevelop} is organised by means of GNU Build System
{\tt Autotools}, which consists of the
well known and widely used set of tools: {\tt  Autoconf+Automake+Libtool}%
\footnote{%
 For more details see Free Software Foundation (FSF) webpages:
 for Automake, Autoconf and Libtool GNU Projects see \url{http://www.gnu.org/software/{automake, autoconf, libtool}}, respectively.
}.
These tools are configured using a user defined
{\tt configure.in} file in the main directory
and {\tt Makefile.am} files in the main directory and all subdirectories.

The {\tt Automake} utility uses all the {\tt Makefile.am} files
 of the project, that are listed in the {\tt configure.in} script
and translates them into {\tt Makefiles}.
The standard {\tt make} utility uses resulting {\tt Makefiles} in order to
compile all relevant source code
and link resulting binaries with the libraries of the project and other shared libraries.
Final executables are located in the relevant subdirectories
created by the build system, see below.

The structure of the project is encoded in the 
{\tt configure.in} and  {\tt Makefile.am} files.
Once it is changed (eg. through adding or removing a new source code, library or directory),
the user may need to edit {\bf both} configuration scripts
( {\tt Makefile.am} and {\tt configure.in}).
Editing them is, however, much less work then creating and maintaining
``manually'' the whole system of
interrelated {\tt Makefile} files in several directories.

The script {\tt configure.in} contains list of all {\tt Makefile.am} 
files used within the project in the macro:
\\ \\
{\tt
AC\_OUTPUT([Makefile MCdev/Makefile Template0/Makefile $\setminus$ \\
Template0/work/Makefile Template/Makefile Template/work/Makefile])
}
\\ \\
This list should be appended with any new {\tt Makefile.am}
in the directory system.
Moreover, every {\tt Makefile.am} specifies subdirectories
(if any) of the project.
For instance {\tt Makefile.am} in the main directory in the current version of 
{\tt MCdevelop} includes the line:\\
\\
{\tt SUBDIRS = MCdev Template0 Template
}
\\ \\
Adding new project directory should be reflected in the above command.
If the new project consists of  subdirectories,
then all files {\tt Makefile.am}
in upper-level directories should list all subdirectories of the lower level
in the directory tree.

According to general philosophy of {\tt Autotools},
all source files, as well as the resulting libraries and binaries are 
specified in files {\tt Makefile.am} in each project's (sub)directories.
Here, the user is referred do manuals of {\tt Autotools}%
\footnote{See \url{http://sources.redhat.com/automake/automake.html\#amhello-Explained}.}.

One may also use {\tt Automake Manager} of {\tt KDevelop}
to create and edit all new {\tt Makefile.am} files of a new project.
On the other hand, there some parts for the new {\tt Makefile.am} files
which has to be added/edited ``by hand'' using text editor.
For example, the {\tt work/Makefile.am } file includes a sizeable
part of the {\tt make} utility instructions,
which are managing multiple batch job preparation and submission
on a PC-farm.
This part of the new {\tt work/Makefile.am } file 
should be copied from our examples and customised slightly
for any any new project,
if the user is interested in running massive parallel jobs on PC-farm.
In particular. in this part of 
the new {\tt work/Makefile.am } one may need to correct
path to the directory {\tt farming}, where  all
C++ scripts described in Section~\ref{sect:farm} reside.

\subsection{Configuring for the use of ROOT package}

Since {\tt MCdevelop} uses ROOT, certain paths and environmental variables
have to be adjusted. It is done automatically by {\tt Autotools} with help of scripts located in {\tt m4} subdirectory. 
Presently, {\tt m4/root.m4} macro is interpreted by {\tt Autoconf}
in order to check the existence of ROOT in the system
and to set up ROOT-related environmental variables (paths to ROOT headers and libraries)
so that the shared libraries of ROOT can be properly used all over the entire project..

Optionally, scripts in {\tt m4 } subdirectory can be adjusted
by the user for linking any other external libraries.

In order to profit from the persistency mechanism of ROOT, it is also necessary
to link the project's classes with the automatic input/output {\em streamer classes}.
Only classes explicitly listed in  the {\tt LinkDef.h} files will be supplied
with automatic streamers and will gain persistency capabilities.
Automatic input/output streamers are produced by the CINT utility of the ROOT
using header files of the classes listed in  {\tt LinkDef.h}
(see \cite{Brun:1997pa} for details).
Adding a new class for which user wishes 
to be supplemented with the streaming functionality
requires adding by hand the following single new line in {\tt LinkDef.h}:
\\
\\
\hbox{~~~~} {\tt \#pragma link C++ class <NewClassName>+;}

\section{Using {\tt KDevelop} IDE}
\label{sec:KDevelop}

The following example shows the usage of the
{\tt KDevelop}, the IDE of KDE
and of a little bit more advanced example program
analysing MC results with the help of ROOT graphics.
Generally, {\tt KDevelop} makes it easier to edit/develop source code
of the project, compile and link it, run and debug executables
and manage {\tt Makefile.am} files of {\tt Automake}.
It makes work of the programmer more comfortable and more efficient.

Note that {\tt KDevelop} is really a graphical front-end
of the {\tt Automake+Autotools} system.
In fact we do not treat this as a disadvantage, 
but rather as an advantage!
This allows any project developed under  {\tt KDevelop} and {\tt MCdevelop}
to be exported into another system (PC-farm) and to be compiled/built/run
from the command line or a shell script without {\tt KDevelop}
being installed there.

\begin{figure}[!t]
\centering
\includegraphics[width=7cm]{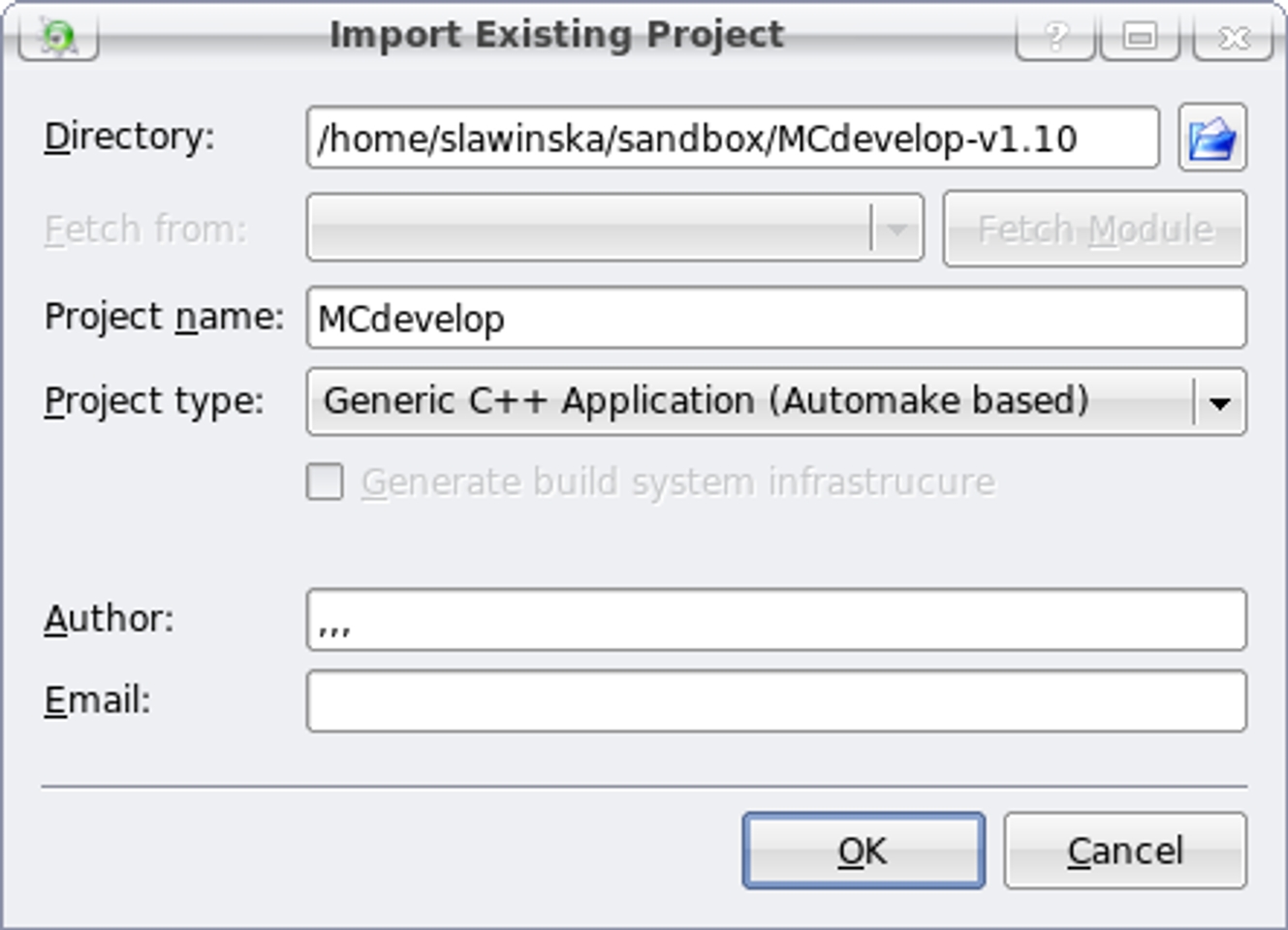}
\caption{Options to be chosen at initial import of a project 
 into {\tt KDevelop}.
 Project type must be set as in the above panel.}
\label{fig:Import}
\end{figure}

\subsection{Importing the Project into {\tt KDevelop}}

Second template example as well as developing the user's 
own project can be done efficiently after importing 
{\tt MCdevelop} into an IDE. In this section we 
present, how to do it on the example of {\tt KDevelop}.
The user familiar with {\tt KDevelop} or using other IDE
can skip this section.

One should start by invoking {\tt KDevelop} from the command line:
\\
\\
{\tt
\$ cd \$MCDEVPATH\\
\$ kdevelop
}
\\
\\
This opens window of the {\tt KDevelop} IDE.
Then, from the upper menu choose: {\tt Project} $\rightarrow$ 
{\tt Import Existing Project...}.
Once a panel visualised on the Figure~\ref{fig:Import}
pops up on the screen,
write the correct {\tt \$MCDEVPATH} path in the top line
(or use the prompt to find it).
Next, it is quite important to choose correct project type.
It must be set as {\em Generic C++ Application (Automake-based)}.

One may also need to choose
{\tt Project} $\rightarrow$  {\tt Build Configuration} 
and tick {\it default} instead of the original {\it debug}
in a pop-up list,
in order to be able to compile and run the program.

\subsection{Using {\tt KDevelop} for compiling and running the Project}

Once the project is imported into {\tt KDevelop},
we can proceed to a more advanced 
template example of the user project
in the distribution directory
and compile/link/run it within this environment.
The following set of commands: 
\\ \\
{\tt \$ cd \$MCDEVPATH/Template}\\
{\tt \$ make install}\\
{\tt \$ cd /work}\\
{\tt \$ root -b -q -l ./Start.C} \\
\\
should be executed in the console of {\tt KDevelop} or other system console.
These commands install the library and to execute an initialisation script
{\tt Start.C}.

\begin{minipage}{6.5cm}
The following step is to open the {\tt Automake Manager}
from the right-hand-side menu of {\tt KDevelop}. 
In the upper frame choose {\tt Template}. \\
\\
The lower frame should now present {\tt libTemplate} with 
implementation of classes building this library and two executable programs: 
{\tt MainPr} and {\tt XPlot}, as shown in the picture on the right.\\
\\
Click on  {\tt MainPr}, then choose a rocket icon
from above the frame in order to compile it and build. 
Then click on the blue cog next to the rocket to execute the program.
\\
\end{minipage}
\begin{minipage}{8.5cm}
\centering
\includegraphics[width=4.5cm, height=10cm]{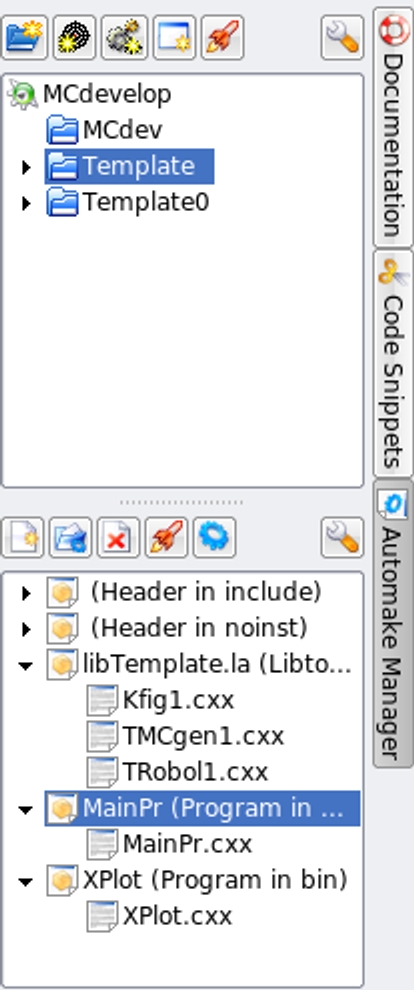}
\end{minipage}
\\ 

Due to a bug in {\tt KDevelop}, one may sometimes need to set the correct 
path for working directory of the executables. 
To do it click on {\tt MainPr (Program in bin)} 
in the {\tt Automake Manager} to highlight it.
Then click on {\tt Options} icon in the {\tt Automake Manager}. 
From the {\it Target Options} for {\tt MainPr} 
window choose the bookmark {\em Argument}.
Then set the working directory path (the third line) to:\\
\\
{\tt \$MCDEVPATH/Template/work}\\
\\
and click the {\em OK} button.
Then repeat the above chain of the commands.

\section{Future developments}

The following future developments of the {\tt MCdevelop} SS software
development environment will have the highest priority:
\begin{itemize}
\item 
adding example demonstrating the possibility of restarting MC production,
\item
adding functionality related to running multiple parallel jobs within
{\tt Grid} and/or {\tt Cloud} type PC-farms,
\end{itemize}

\vspace{4mm}
\noindent
{\bf\Large Acknowledgements}\\
We would like to thank M. Skrzypek, W. Placzek and Z. Was, 
the co-authors of previous stochastic simulation
project published by the Krakow group.
Their ideas and experience were naturally incorporated into {\tt MCdevelop}. 
We would like to acknowledge P. Golonka for his help in using {\tt Autotools} 
with ROOT. We thank M. Skrzypek, W. Placzek, P. Richardson and D. Grellscheid for
useful comments and reading the manuscript.


\begingroup	\begin{thebibliography}{1}

\bibitem{bhlumi4:1996}
S.~Jadach, W.~P\l{}aczek, E.~Richter-Was, B.~F.~L. Ward, and Z.~Was, {\em
  Comput. Phys. Commun.} {\bf 102} (1997)
229.

\bibitem{Jadach:2007qa}
S.~Jadach, W.~P\l{}aczek, M.~Skrzypek, P.~Stephens, and Z.~Was,
\href{http://www.arXiv.org/abs/hep-ph/0703281}{{\tt hep-ph/0703281}}.

\bibitem{Jadach:2008nu}
S.~Jadach, W.~Placzek, M.~Skrzypek, and P.~Stoklosa, {\em Comput. Phys.
  Commun.} {\bf 181} (2010) 393--412,
\href{http://www.arXiv.org/abs/0812.3299}{{\tt 0812.3299}}.

\bibitem{Brun:1997pa}
R.~Brun and F.~Rademakers, {\em Nucl. Instrum. Meth.} {\bf A389} (1997)
81--86.

\bibitem{foam:2002}
S.~Jadach, {\em Comput. Phys. Commun.} {\bf 152} (2003) 55--100,
\href{http://www.arXiv.org/abs/physics/0203033}{{\tt physics/0203033}}.

\end{thebibliography}\endgroup
\providecommand{\href}[2]{#2}
\end{document}